\documentclass[journal=jacsat,manuscript=article,layout=twocolumn]{achemso}

\usepackage{amsmath}
\usepackage{amssymb}
\usepackage{graphicx}% Include figure files
\usepackage{dcolumn}% Align table columns on decimal point
\usepackage{bm}% bold math
\usepackage{xcolor}
\usepackage{braket}
\usepackage[version=3]{mhchem}
\usepackage{xfrac}
\usepackage{float}
\usepackage{afterpage}
\title{Free-spin dominated magnetocaloric effect in dense Gd$^{3+}$ double perovskites}

\author{EliseAnne C. Koskelo}
\altaffiliation{Now at Department of Physics, Harvard University, Cambridge, MA 02138, USA.}
\author{Cheng Liu}
\author{Paromita Mukherjee}
\author{Nicola D. Kelly}
\author{Si\^{a}n E. Dutton}
\email{sed33@cam.ac.uk}
\affiliation{
Department of Physics, University of Cambridge, Cambridge CB3 0HE, United Kingdom }

\date{\today}% It is always \today, today,
\begin{document}

\begin{abstract}
Frustrated lanthanide oxides with dense magnetic lattices are of fundamental interest for their potential in cryogenic refrigeration due to a large ground state entropy and suppressed ordering temperatures, but can often be limited by short-range correlations. Here, we present examples of frustrated \emph{fcc} oxides, Ba$_2$GdSbO$_6$ and Sr$_2$GdSbO$_6$ and the new site-disordered analog Ca$_2$GdSbO$_6$ ([CaGd]$_A$[CaSb]$_B$O$_6$), in which the magnetocaloric effect is influenced by minimal superexchange ($J_1 \sim 10$ mK). We report on the crystal structures using powder x-ray diffraction and the bulk magnetic properties through low-field susceptibility and isothermal magnetization measurements. The Gd compounds exhibit a magnetic entropy change of up to -15.8 J/K/mol$_\textrm{Gd}$ in a field of 7 T at 2 K, a 20\% excess compared to the value of -13.0 J/K/mol$_\textrm{Gd}$ for a standard in magnetic refrigeration, Gd$_3$Ga$_5$O$_{12}$. Heat capacity measurements indicate a lack of magnetic ordering down to 0.4 K for Ba$_2$GdSbO$_6$ and Sr$_2$GdSbO$_6$, suggesting cooling down through the liquid 4-He regime. A mean-field model is used to elucidate the role of primarily free spin behavior in the magnetocaloric performance of these compounds in comparison to other top-performing Gd-based oxides. The chemical flexibility of the double perovskites raises the possibility of further enhancement of the magnetocaloric effect in the Gd$^{3+}$ \emph{fcc} lattices.
%\begin{description}
%\item[Keywords] 
%Rare earth oxide, magnetocaloric, adiabatic demagnetization, magnetic correlations
%\end{description}
\end{abstract}

\maketitle

%\tableofcontents

\section{Introduction}

Cryogenic cooling is imperative to modern technologies including quantum computing and magnetic resonance imaging. While liquid He can be used to reach temperatures as low as 20 mK (using 3-He and 4-He) or 2 K (4-He only), it is a depleting resource and sustainable alternatives capitalizing on magnetic, structural, and/or electric ordering of materials are of key interest.\cite{Science_Moya_Caloric} In adiabatic magnetic refrigerators (ADRs), an applied magnetic field induces a change in entropy of the spins of a material, $\Delta S_m$. 
When followed by adiabatic demagnetization, the system exhibits a proportional decrease in temperature $\Delta T$ as dictated by the magnetocaloric (MC) effect. MC materials are operable at temperatures above their ordering transition $T_0$, and are
often characterized by the maximum isothermal magnetic entropy change $\Delta S_m$ that can be achieved for a given change in field $\Delta H$.\cite{JAP_MCE_indirect} Current top-performing MC materials are based on Gd$^{3+}$ containing compounds as the minimal single-ion anisotropy ($L = 0$) of the magnetic ions allows for full extraction of the theoretical entropy change in high magnetic fields.\cite{Gd-formate_2013,GdPO4_MCE_2014,GdF3_MCE_2015} Recent advances in materials like Gd(HCOO)$_3$, GdF$_3$, and GdPO$_4$, with dense magnetic sublattices, have highlighted the importance of weak magnetic correlations in enabling a large MC effect.\cite{Gd-formate_2013,GdPO4_MCE_2014,MCE_review_2020} However, these materials can be limited by a lack of tunability via chemical substitution and/or large volumes per magnetic ion. \cite{GdF3_MCE_2015,MCE_review_2020} %For example, GdVO$_4$ was investigated recently, and while comparable to GdPO$_4$ in its entropy change at finite fields, it has a relatively large ordering temperature of 2.5 K due to strong exchange interactions.\cite{GdVO4_2018} 
\textcolor{black}{This limits the opportunities for tuning the magnitude and temperature of the maximum $\Delta S_m$.}

On the other hand, frustrated magnetic oxides, in which the geometry of the lattice prevents all exchange interactions from being satisfied simultaneously, present a diverse class of MC candidates given their chemical stability and exotic magnetic properties such as a large ground state degeneracy and suppressed ordering temperature.\cite{Zhitomirsky_2003} Gd$_3$Ga$_5$O$_{12}$, a frustrated garnet, is the standard among this class of materials, with an entropy change of -13.0 J/K/mol$_\textrm{Gd}$ in a field of 7 T at 2 K.\cite{GdF3_MCE_2015} In addition to the garnet lattice which is comprised of two interpenetrating networks of bifurcating loops of ten corner-sharing $Ln^{3+}$ triangles, a wealth of frustrated geometries exist, including the pyrochlore and kagome lattices, and the $fcc$ lattice, which as a set of edge-sharing tetrahedra is frustrated under antiferromagnetic exchange. However, one limitation of oxide materials is that their MC performance can be strongly influenced by short-range correlations.\cite{Gd2Ti2O7_MCE} For example, Gd$_3$Ga$_5$O$_{12}$ has a relatively large superexchange between Gd$^{3+}$ ions, $|J_1|\sim 100$ mK \cite{Paddison_GGG_2015}, compared to $|J_1| \sim 70$ mK for GdF$_3$ \cite{Gd-formate_2013}.

Lanthanide oxide double perovskites with the general formula $A_2Ln$SbO$_6$ ($A$ = alkaline earth (Ba, Sr), $Ln$ = lanthanide) represent a family of frustrated magnets in which $Ln$ ions lie on an \emph{fcc} magnetic sublattice, enforced by the rock-salt arrangement with the other $B$ site cation, Sb$^{5+}$. Chemical pressure via $A$ site cation substitution can alter the distortion of a single $Ln$-ion tetrahedra due to small changes in the nearest neighbor $nn$ distance dictated by rotations of the $B$O$_6$ octahedra \cite{Karunadasa_PNAS}.

Here we present $fcc$ oxides with minimal superexchange, up to 10 times smaller than other frustrated oxides. We report on the solid-state synthesis, structural characterization, bulk magnetic properties, and magnetocaloric effect in these materials, Ba$_2$GdSbO$_6$ and Sr$_2$GdSbO$_6$, and the new site-disordered Ca analog, Ca$_2$GdSbO$_6$ ([CaGd]$_A$[CaSb]$_B$O$_6$). We show that tuning of the $A$ site ion influences the exchange through changes in the nearest neighbor distance and O-Gd-O bond angles, with small effect on the overall magnetocaloric performance, suggesting that free spin behavior is dominant for 1.8 K and above. Furthermore, the low Curie-Weiss temperatures ($\sim$0.8 K), frustrated lattice geometry, and lack of ordering of the Ba and Sr compounds suggest that cooling may persist through the cooperative paramagnetic regime to temperatures well below 0.4 K. %The related compound, Sr$_2$GdNbO$_6$, was recently reported, but exhibits fundamentally different magnetic properties (e.g. ferromagnetic superexchange) \cite{Sr2GdNbO6_MCE}, which could be attributable to $d^0$ versus $d^{10}$ effects found in other lanthanide oxides (need some suggested citations here). The compound exhibits a large MC effect of -15.5 J/K/mol$_\textrm{Gd}$ in a field of 7 T, but is maximized at 3 K, before dropping to -10.9 J/K/mol$_\textrm{Gd}$ at 2 K \cite{Sr2GdNbO6_MCE}.

\begin{figure*}[!ht]
    \centering
    \includegraphics[width=1.8\columnwidth,clip,trim={120 20 260 70}]{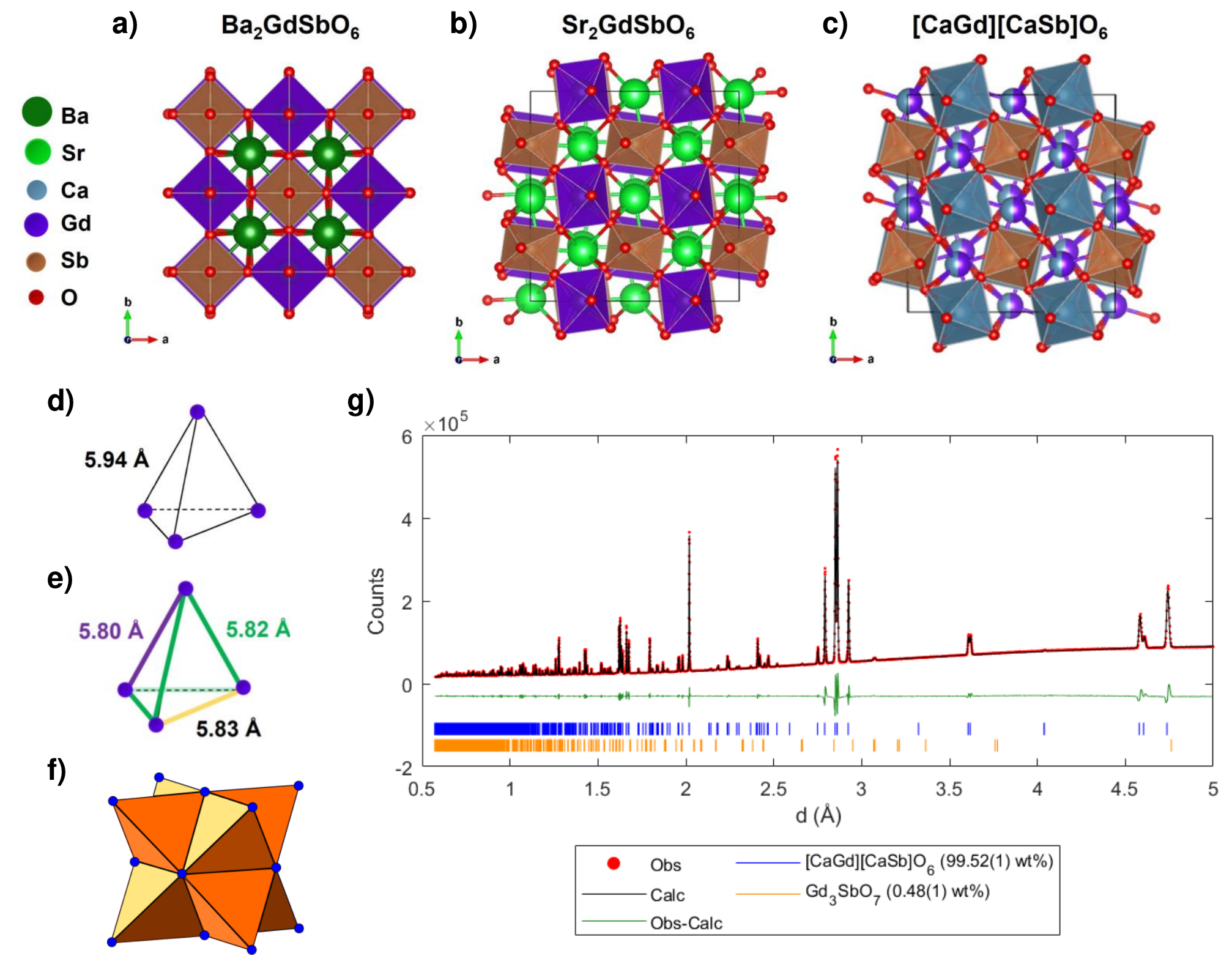}
    \caption{Crystal structures of the double perovskites a) Ba$_2$GdSbO$_6$, b) Sr$_2$GdSbO$_6$ and c) [CaGd]$_A$[CaSb]$_B$O$_6$. In [CaGd]$_A$[CaSb]$_B$O$_6$, the Gd$^{3+}$ ions lie in a disordered arrangement with Ca$^{2+}$ on the $A$ sites. The rock-salt ordering of Gd$^{3+}$ and Sb$^{5+}$ on the $B$ sites produces a \emph{fcc} magnetic sublattice which is d) uniform for Ba$_2$GdSbO$_6$ and e) distorted for Sr$_2$GdSbO$_6$, with the listed side lengths. f) The \emph{fcc} sublattice is a frustrated geometry because it composes a network of edge-sharing tetrahedra. g) High-resolution powder x-ray diffraction Rietveld refinement of [CaGd]$_A$[CaSb]$_B$O$_6$. Observed intensities and calculated intensities obtained from a Rietveld refinement are shown as red circles and a black line, respectively; the difference (data–fit) is shown by a green line. Reflection positions are indicated by blue and orange tick marks, for phases [CaGd]$_A$[CaSb]$_B$O$_6$ and the small phase impurity Gd$_3$SbO$_7$ (0.48(1)\% by weight), respectively. }
    \label{fig:crystals}
\end{figure*}

\section{Experimental Methods}
\label{section:methods}

\subsubsection{Solid State Synthesis}
Powder samples, \textcolor{black}{ of $\sim$1 g}, of Ba$_2$GdSbO$_6$, Sr$_2$GdSbO$_6$, and Ca$_2$GdSbO$_6$ were prepared as described in the literature \cite{Karunadasa_PNAS}. Stoichiometric mixtures of pre-dried gadolinium (III) oxide (99.999\%, Alfa Aesar REacton), antimony (V) oxide (99.9998\%, Alfa Aesar Puratronic), and the appropriate alkaline earth carbonate, barium carbonate (99.997\%, Alfa Aesar Puratronic), strontium carbonate (99.99\%, Alfa Aesar), or calcium carbonate (99.99\%, Alfa Aesar Puratronic), were initially ground using a mortar and pestle and heated in air at 1400\textdegree C for 24 hours. Heating was repeated until the amount of impurity phases as determined by x-ray diffraction no longer reduced upon heating (one additional 24 hour cycle). Ba$_2$GdSbO$_6$, Sr$_2$GdSbO$_6$, \textcolor{black}{and Ca$_2$GdSbO$_6$} contain an impurity phase of Gd$_3$SbO$_7$ of 0.45(4) wt\%, 0.67(4) wt\%, \textcolor{black}{and 0.48(1) wt\%, respectively}, which orders antiferromagnetically at 2.6 K \cite{Hinatsu_Gd3SbO7}. %Stoichiometric amounts of each alkaline carbonate were mixed with Gd$_2$O$_3$ and Sb$_2$O$_5$ and ground with a mortar and pestle. Mixtures were fired at 1400 \textdegree C for 2 to 4 24-hr periods. Phase purity was monitored using x-ray powder diffraction.

\subsubsection{Structural Characterization}

Room temperature powder x-ray diffraction (XRD) measurements were carried out using a Bruker D8 Advance diffractometer (Cu K$\alpha$ radiation, $\lambda$ = 1.54 \AA). Data was collected with $d(2\theta)=0.01$\textdegree \hspace{1pt} from $2\theta$=15-150\textdegree \hspace{1pt}, with an overall collection time of 2-3 hours. During each scan, the sample stage was rotated to avoid preferred orientation effects. Additional high resolution x-ray powder diffraction measurements were conducted at the I11 beamline at the Diamond Light Source using a position sensitive detector for Ca$_2$GdSbO$_6$ at room temperature and at 100 K. \textcolor{black}{Data was collected with $\lambda$ = 0.826866 \AA \hspace{1pt} from $2\theta$= 8-100\textdegree \hspace{1pt} using a position sensitive detector, with an overall collection time of one minute.} The powder sample was mounted in a 0.28 mm diameter capillary inside a brass sample holder.

Rietveld refinements \cite{McCusker_RietveldRefinement} of the powder XRD data were completed using the Diffrac.Suite TOPAS5 program \cite{TOPAS_Academic}. Peak shapes were modeled using a pseudo-Voigt function \cite{TCHZ_peaks} and the background was fit using a 13-term Chebyshev polynomial. Except for Ca$_2$GdSbO$_6$, \textcolor{black}{where synchrotron XRD data was available,}  all Debye-Waller factors were kept constant to the literature reported values from powder neutron diffraction \cite{Karunadasa_PNAS,wong-ng_x-ray_2014}. A cylindrical correction was used to correct for capillary absorption in the I11 data as well as a Lorentzian/Gaussian model to account for strain broadening effects on the peak shape \cite{balzar_sizestrain_2004}.

\subsubsection{Magnetic Characterization}

Magnetic susceptibility $\chi(T) = dM/dH$ ($\sim M/H$ in the low field limit) and isothermal magnetization $M(H)$ measurements were conducted using a Quantum Design Magnetic Properties Measurement System (MPMS) with a superconducting interference device (SQUID) magnetometer. Susceptibility measurements were made in zero-field-cooled conditions ($\mu_0 H$=1000 Oe, where $M(H)$ is linear and the $\chi = M/H$ approximation is valid) over a temperature range of 1.8-300 K and in field-cooled conditions from $T=1.8$-30 K. $M(H)$ measurements were made over a field range of $0 \leq \mu_0 H \leq 7$ T, in steps of 0.2 T from 2 to 20 K, in 2 K steps \textcolor{black}{from 2 to 10 K and in 5 K steps from 10 to 20 K}. The magnetic entropy change for a field $H_{max}$ relative to zero field, was extracted from $M(H)$ by computing the temperature derivative of the magnetization using:
\textcolor{black}{
\begin{equation}
   \left( \frac{\partial M (T_0, H)}{\partial T} \right)_H \approx \frac{M(T_{i+1},H)-M(T_i,H)}{T_{i+1} - T_i},
   \label{eq:finiteDiff}
\end{equation}
and then integrating in discrete steps of 0.1 T across fields using the trapezoidal method:
\begin{equation}
    \Delta S_m(T_0, H_{max}) = \int_0^{H_{max}} \left(\frac{\partial M (T_0,H)}{\partial T} \right)_H dH.
    \label{eq:deltaSm_fromdMdT}
\end{equation}}

\subsubsection{Low Temperature Heat Capacity}
Magnetic heat capacity measurements were carried out using a Quantum Design PPMS using the 3-He probe (0.4$\leq$T$\leq$30 K) \textcolor{black}{in zero field}. Equal masses of the sample and Ag powders were mixed with a mortar and pestle and pressed into a 0.5-mm pellet to enhance the thermal conductivity. Pellets were mounted onto the sample platform using $N$-grease to ensure thermal contact. Addenda measurements of the sample platform and grease were calibrated at each temperature before measurement. The sample heat capacity, $C_p$, was obtained from the measured heat capacity, $C_{tot}$, by subtracting the Ag-contribution using the literature values \cite{Ag_low-temperature_1995}. The magnetic heat capacity $C_{mag}$ was obtained from a subtraction of the lattice contribution $C_{lat}$ from the sample heat capacity $C_p$:
\begin{equation}
C_{mag}(T,H) = C_p(T,H) - C_{lat}(T,H).
\label{eq :Cm_fromCp}
\end{equation}
The lattice contribution $C_{lat}$ was determined using least-squares fits of the zero-field $C_p$ at high temperatures (8-50 K) to the Debye model:
\begin{equation}
C_{lat}(T) = \frac{9nRT^3}{T_D^3} \int_0^{T_D/T} \frac{x^4 e^x}{(e^x-1)^2}dx,
\label{eq:DebyeModel}
\end{equation}
where $T_D$ is the Debye temperature, $R$ is the molar gas constant, and $n$ is the number of atoms per formula unit. The total magnetic entropy, relative to the lowest temperature $T_i$ measured, was computed using:
\begin{equation}
S_m(T,H) = \int_{T_i}^{T} \frac{C_{mag}(T’,H)}{T’} dT’,
\label{eq:magEntropy_fromHC}
\end{equation}
\textcolor{black}{numerically for the temperatures measured.}

\section{Structural Characterization}

Powder x-ray diffraction indicate formation of almost phase pure sample for $A_2$GdSbO$_6$ ($A$ = \{Ba, Sr, Ca\}). Rietveld refinements, Figure \ref{fig:crystals} and Table \ref{tab:Rietveld}, show that all three compounds exhibit small $<$0.7 wt\% impurities of Gd$_3$SbO$_7$ \cite{Hinatsu_Gd3SbO7}. The structures of Ba$_2$GdSbO$_6$ and Sr$_2$GdSbO$_6$ are consistent with prior reports \cite{Karunadasa_PNAS}. Both materials exhibit full rock-salt ordering of Gd$^{3+}$ and Sb$^{5+}$ on the $B$ sites, attributable to the large charge and ionic radii differences between cations; Ba$_2$GdSbO$_6$ forms a cubic $Fm\bar{3}m$ structure, resulting in a uniform tetrahedron of Gd$^{3+}$ ions, while Sr$_2$GdSbO$_6$ forms a monoclinic $P2_1/n$ structure, resulting in a distorted tetrahedron of Gd$^{3+}$ ions, Figure \ref{fig:crystals} \cite{Karunadasa_PNAS}. 

\begin{table*}[ht!]
\caption{\label{tab:Rietveld} Lattice parameters and crystal structures of Ba$_2$GdSbO$_6$, Sr$_2$GdSbO$_6$, and Ca$_2$GdSbO$_6$ ([CaGd]$_A$[CaSb]$_B$O$_6$) as determined from Rietveld refinements of powder x-ray diffraction at room temperature for Ba$_2$GdSbO$_6$ and Sr$_2$GdSbO$_6$ and at 100 K for [CaGd]$_A$[CaSb]$_B$O$_6$. The Debye-Waller factors for Ba$_2$GdSbO$_6$ and Sr$_2$GdSbO$_6$ were kept constant to values reported in the literature for the related compound Ba$_2$DySbO$_6$ and Sr$_2$GdSbO$_6$, respectively.\cite{Karunadasa_PNAS,wong-ng_x-ray_2014}}
%\begin{ruledtabular}
\begin{tabular}{lcccccc}
\hline
\hline
Material & Atom & Wyckoff position & $x$ & $y$ & $z$ & $\textrm{B}_{\textrm{iso}} (\AA^2)$  \\  \hline

\textbf{Ba$_2$GdSbO$_6$} & Ba & $8c$ & 0.25 & 0.25 & 0.25 & 0.67\\
\hspace{6pt} $Fm\bar{3}m$ & Gd & $4a$ & 0 & 0 & 0 & 0.48 \\
& Sb & $4b$ & 0.5 & 0.5 & 0.5 & 0.41 \\
& O & $24e$ & 0.257(2) & 0 & 0 & 0.76 \\
\hspace{6pt} $a (\AA)$ & & 8.47517(2) & & & & \\
\hspace{6pt} Gd$_3$SbO$_7$ (wt\%) & & 0.45(4) & & & & \\
\hspace{6pt} $\chi^2$ & & 1.41 & & & & \\
\hspace{6pt} $R_{wp}$ & & 11.6 & & & & \\
\hline
\textbf{Sr$_2$GdSbO$_6$} & Sr & $4e$ & 0.0105(5) & 0.0346(2) & 0.2489(7) & 0.79 \\
\hspace{6pt} $P2_1/n$ & Gd & $2d$ & 0.5 & 0 & 0 & 0.24 \\
& Sb   & $2c$ & 0 & 0.5 & 0 & 0.39 \\
& O(1) & $4e$ & 0.253(3) & 0.317(3) & 0.021(3) & 0.79 \\
& O(2) & $4e$ & 0.190(4) & 0.761(3) & 0.042(3) & 0.79 \\
& O(3) & $4e$ & -0.086(2) & 0.486(2) & 0.239(3) & 0.79 \\
\hspace{6pt} $a$ (\AA) & & 5.84113(5) & & & & \\
\hspace{6pt} $b$ (\AA) & & 5.89402(5) & & & & \\
\hspace{6pt} $c$ (\AA) & & 8.29127(7) & & & & \\
\hspace{6pt} $\beta$ (\textdegree) & & 90.2373(7) & & & & \\
\hspace{6pt} Gd$_3$SbO$_7$ (wt\%) & & 0.67(4) & & & &\\
\hspace{6pt} $\chi^2$ & & 1.20 & & & & \\
\hspace{6pt} $R_{wp}$ & & 8.7 & & & & \\
\hline

\textbf{[CaGd]$_A$[CaSb]$_B$O$_6$} & Ca$_1$/Gd & $4e$ & -0.0174(1) & 0.05939(7) & 0.25403(8) & 0.90(1) \\ %I11 refinement
%P2$_1$/n & Occ$_{\textrm{Ca}_1}$ & 0.5 & & & &  \\
 %& Occ$_{\textrm{Gd}}$ & 0.5 & & & &  \\
\hspace{6pt} $P2_1/n$ & Ca$_2$ & $2d$ & 0.5 & 0 & 0 & 0.68(8) \\
& Sb & $2c$ & 0 & 0.5 & 0 & 0.54(1) \\
& O(1) & $4e$ & 0.1660(8) & 0.2149(7) & -0.0728(6) & 1.25(5) \\
& O(2) & $4e$ & 0.2089(8) & 0.1769(7) & 0.5511(6) & 1.25(5) \\
& O(3) & $4e$ & 1.1205(7) & 0.4390(7) & 0.2254(5) & 1.25(5) \\
\hspace{6pt} $a$ (\AA) & & 5.58025(2) & & & & \\
\hspace{6pt} $b$ (\AA) & & 5.84820(2) & & & & \\
\hspace{6pt} $c$ (\AA) & & 8.07706(2) & & & & \\
\hspace{6pt} $\beta$ (\textdegree) & & 90.3253(2) & & & & \\
\hspace{6pt} Occ$_{\textrm{Ca}_1}$ & & 0.5 & & & &  \\
\hspace{6pt} Occ$_{\textrm{Gd}}$ & & 0.5 & & & &  \\
\hspace{6pt} Gd$_3$SbO$_7$ (wt\%) & & 0.48(1) & & & &\\
\hspace{6pt} $\chi^2$ & & 5.80 & & & & \\
\hspace{6pt} $R_{wp}$ & & 3.7 & & & & \\
\hline
\hline
\end{tabular}
%\end{ruledtabular}
\end{table*}

Structural refinements, Table \ref{tab:Rietveld}, indicate that Ca$_2$GdSbO$_6$ adopts the monoclinic space group $P2_1/n$. However, additional reflections ((0 1 1) and (1 0 1)) at $2\theta \approx 18$\textdegree ($d = 4.6-4.8$ \AA) indicate that \textcolor{black}{Gd$^{3+}$ occupies the $A$ sites as in the case of its of nonmagnetic analog [CaLa]$_A$[CaSb]$_B$O$_6$ \cite{Ca2LaSbO6_Asite_mixing}. Refinement of the occupancy of Gd$^{3+}$ and Ca$^{2+}$ across the $A$ and $B$ sites indicated that Gd$^{3+}$ only occupies the $A$ sites and so this cation distribution was fixed in subsequent refinements.}  Thus a clearer description of the compound is [CaGd]$_A$[CaSb]$_B$O$_6$.\footnote{Here, the braketed notation $[XY]_A[LZ]_B$O$_6$ refers to a double perovskite in which species $X$ and $Y$ lie on the $A$ sites, and species $L$ and $Z$ lie on the $B$ sites.} Consistent with a prior study of Mn-doped [Ca$_{1-x}$Sr$_x$Gd]$_A$[CaSb$_{1-y}$]$_B$O$_6$:Mn$_y$ ($x$ = 0.4, $y$ = 0.003) \cite{Ca2GdSbO6_doped} and the structure of [CaLa]$_A$[CaSb]$_B$O$_6$ \cite{Ca2LaSbO6_Asite_mixing}, we find a disordered arrangement on the $A$ sites with half of the sites occupied by the Gd$^{3+}$ ions and the other half by Ca$^{2+}$, and rock-salt ordering of Ca$^{2+}$ and Sb$^{5}$ on the $B$ sites. 

There is no evidence of $A$-site ordering in the compound with no further superstructure peaks observed in the synchrotron XRD. This is likely due to the minimal charge and ionic radii differences of the Ca$^{2+}$ and Gd$^{3+}$ ions which rules out rock-salt and columnar ordering on the $A$ sites \cite{king_cation_2010}.

To analyze whether the site disorder in [CaGd]$_A$[CaSb]$_B$O$_6$ is due to close-packing efficiency considerations, we computed the Goldschmidt tolerance factor (GTF), $t$. $t$ predicts whether the ionic radii of the $A$-site cation and $B$-site cation are well-scaled for a cubic structure in which $A$-site cations lie at the cavities of $B$-O octahedra \cite{goldschmidt_gesetze_1926}. It can be extended to double perovskites with the formula $AA'BB'$O$_6$ by computing an average $A$-site and $B$-site ionic radii so that:
\begin{equation}
    t = \frac{r_{\textrm{A},\textrm{avg}}+r_{\textrm{O}}}{\sqrt{2}(r_{B,\textrm{avg}} + r_{\textrm{O}})},
    \label{eq:GTF}
\end{equation}
where $r_{A,\textrm{avg}} = \frac{1}{2}r_{\textrm{A}} + \frac{1}{2}r_{A'}$ \cite{vasala_a2bbo6_2015}. The GTF for [CaCa]$_A$[GdSb]$_B$O$_6$ in which Gd$^{3+}$ and Sb$^{5+}$ are located on the $B$ sites is 0.82 compared to 0.80 for [CaGd]$_A$[CaSb]$_B$O$_6$ in which Gd$^{3+}$ lies on the $A$ sites. Since $t \approx$1 in stable perovskite structures, these results suggest that the Gd$^{3+}$ ions should lie preferentially on the $B$-site. 

It is also possible that charge differences of the cations stabilize the observed disordered structure \cite{king_cation_2010}. However, an additional calculation using a charge-based tolerance factor also predicted [CaCa]$_A$[GdSb]$_B$O$_6$ as the more stable structure \cite{charge_based_tau}. Thus, the $A$ site occupancy of Gd$^{3+}$ is unlikely to be due to the greater close cubic packing efficiency or charge differences between cations. \textcolor{black}{It could be that the additional entropy associated with a random distribution of Ca$^{2+}$ and Gd$^{3+}$ on the $A$ site favours the observed cation distribution, although we note that a random distribution of Ca$^{2+}$/Gd$^{3+}$ across both sites is more entropically favourable.} Changes to the synthesis procedure e.g. slow cooling may result in differences in the Ca$^{2+}$/Gd$^{3+}$ distribution, \textcolor{black}{but are beyond the scope of this study.}

%\begin{figure*}[htbp]
   % \centering
    %\includegraphics[width=1.9\columnwidth,clip,trim={0 20 0 0}]{refined-Ca2GdSbO6-chiSq5p8-boxed.eps}
  %  \caption{High-resolution powder x-ray diffraction Rietveld refinement of [CaGd]$_A$[CaSb]$_B$O$_6$. Observed intensities and calculated intensities obtained from a Rietveld refinement are shown as red circles and a black line, respectively; the difference (data–fit) is shown by a green line. Reflection positions are indicated by blue and orange tick marks, for phases [CaGd]$_A$[CaSb]$_B$O$_6$ and Gd$_3$SbO$_7$, respectively. }
   % \label{fig:Rietveld_fit}
%\end{figure*}

\section{Magnetic Characterization \& Results}

%\begin{figure}[htbp]
  %  \centering
  %  \includegraphics[width=\columnwidth]{dim-plot-Gd-series-withinset.eps}
   % \caption{Dimensionless inverse magnetic susceptibility of $A_2$GdSbO$_6$ ($A$ = \{Ba, Sr, Ca\}) scaled by the appropriate factors of the Curie constant $C$ and Curie-Weiss temperature $\Theta$ for each material. The error bars are smaller than the points in the graph. All compounds remain paramagnetic down to 1.8 K. The inset depicts the percent deviation of $\chi^{-1}$ from the Curie-Weiss law, $C/(T-\Theta)$. Error bars are determined assuming a 0.1 mg mass error. }
  %  \label{fig:dim-chi}
%\end{figure}

\begin{figure}[htb!]
    \centering
    \includegraphics[width=0.98\columnwidth,clip,trim={70 330 200 60}]{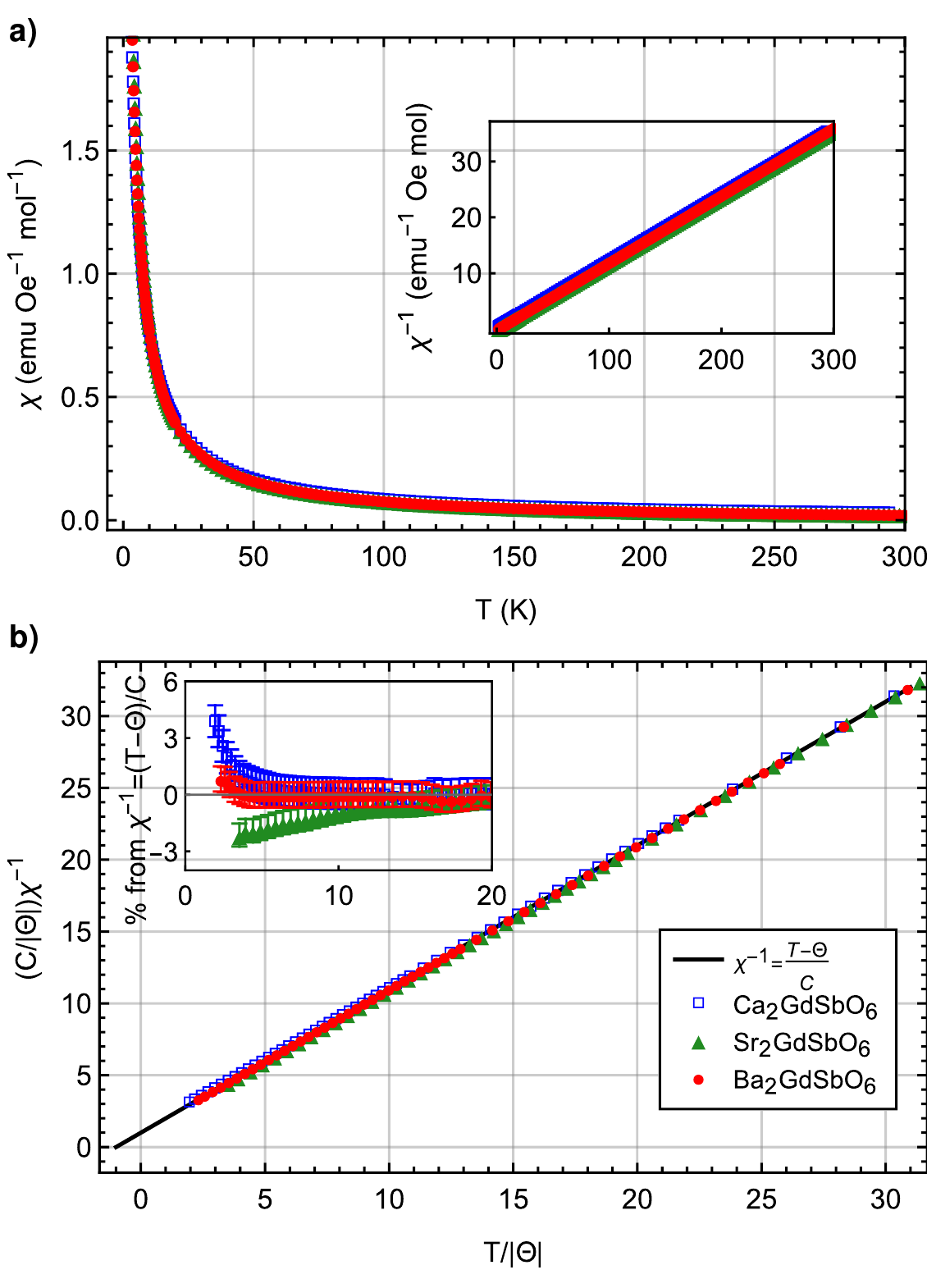}
    \caption{a) Low-field magnetic susceptibility $\chi \approx M/H$ versus temperature $T$ of $A_2$GdSbO$_6$ ($A$ = \{Ba, Sr, Ca\}) (inset: $\chi^{-1}$ versus $T$) b) Dimensionless inverse magnetic susceptibility scaled by the appropriate factors of the Curie constant $C$ and Curie-Weiss temperature $\Theta$ for each material. The error bars are smaller than the points in the graph. All compounds remain paramagnetic down to 1.8 K. The inset depicts the percent deviation of $\chi^{-1}$ from the Curie-Weiss law, $C/(T-\Theta)$. Error bars are determined assuming a 0.1 mg mass error. }
    \label{fig:dim-chi}
\end{figure}

The zero-field-cooled (ZFC) magnetic susceptibility $\chi(T)$, Figure \ref{fig:dim-chi}, indicates paramagnetic behavior of each material down to 1.8 K, in agreement with prior investigations \cite{Karunadasa_PNAS}. Curie-Weiss fits to the inverse susceptibility $\chi^{-1}(T)$ were conducted from temperatures $T$ = 8 to 50 K. For all compositions, the negative Curie-Weiss temperatures $\Theta_{CW}$, Table \ref{table:J1_fits_from_MFT}, indicate antiferromagnetic interactions between spins that increase in strength from the monoclinic ($A$ = Sr) to cubic lattice ($A$ = Ba). The Curie-Weiss law can be written in a dimensionless form given by: \textcolor{black}{$\frac{C}{\chi|\Theta|} = \frac{T}{|\Theta|} + 1$}
(for $\Theta_{CW}<$0), where $\Theta_{CW}$ is the Curie temperature and $C$ is the Curie constant.\cite{Melot_SRO_spinels_2009} This dimensionless form can elucidate the presence of short-range correlations from the inverse magnetic susceptibility and enable a comparison across compounds \cite{Melot_SRO_spinels_2009,Sian_spinels}. In these dimensionless units, free spin behavior is indicated by the linear relationship of \textcolor{black}{$\frac{C}{\chi|\Theta|}$} with \textcolor{black}{$\frac{T}{|\Theta|}$} with an y-intercept of 1. Positive (negative) deviations from linearity can indicate antiferromagnetic (ferromagnetic) short-range correlations between spins. Figure \ref{fig:dim-chi} highlights the minimal short-range correlations of each $A_2$GdSbO$_6$ compound, all of which exhibit extremely small deviations of less than 5\% from free-spin behavior at 1.8 K. As a comparison, the MgCr$_2$O$_4$ spinels, which order at $\frac{T}{|\Theta|} \sim 0.03$, exhibit 20-60\% deviations at temperatures of $\frac{T}{|\Theta|} \sim 0.1$ \cite{Sian_spinels}. The positive deviations of [CaGd]$_A$[CaSb]$_B$O$_6$ are likely to correspond to antiferromagnetic short-range correlations rather than disorder-induced quantum fluctuations. Field-cooled measurements below 20 K indicate no hysteresis (see Figure S1 and S2

\begin{figure}[t!]
    \centering
    \includegraphics[width=\columnwidth,clip,trim={80 280 80 100}]{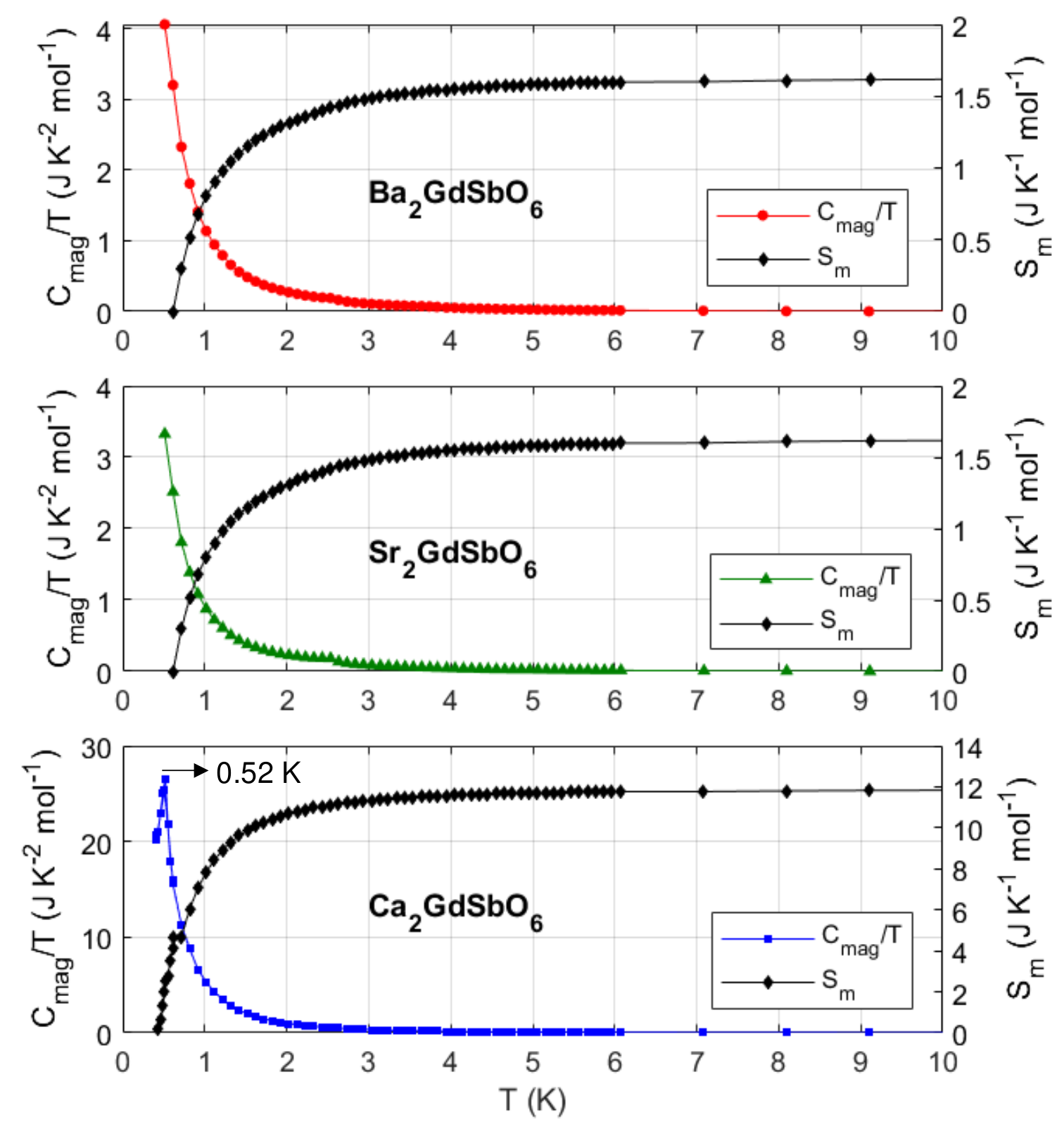}
    \caption{\textcolor{black}{Zero field} magnetic heat capacity normalized by temperature $C_{mag}/T$ of $A_2$GdSbO$_6$ ($A =$\{Ba, Sr, Ca\}), and corresponding magnetic entropy released from 0.4-10 K. The Debye temperatures for each material (Table \ref{table:J1_D_Gd_compounds}), were found from fits of the total heat capacity to the Debye model (Equation \ref{eq:DebyeModel}) from 6-30 K. The small anomaly in each measurement at 2.6 K is likely due to the ordering of the $\sim$0.5-1 wt\% Gd$_3$SbO$_7$ impurity \cite{Hinatsu_Gd3SbO7}.}
    \label{fig:He3_Gd_HC}
\end{figure}

A mean-field estimate for the nearest neighbor ($nn$) superexchange $J_1$, Table \ref{table:J1_fits_from_MFT}, indicates weak coupling between spins in the \emph{fcc} compounds, $\sim$ 10 mK, compared to 100 mK in \ce{Gd3Ga5O_{12}} \cite{Paddison_GGG_2015}. Although the number of nearest neighbors is not constant for the disordered [CaGd]$_A$[CaSb]$_B$ analog, a mean-field estimate for the six nearest $A$ sites ($R_{nn,avg}=4.063(3)$ \AA \hspace{1pt} from distances 2$\times$3.918(3), 2$\times$4.171(3), 4.162(3), and 4.038(3) \AA) was computed, giving an order of magnitude estimate for $J_1$.

The \textcolor{black}{zero field} measured magnetic heat capacities $C_{mag}$ of $A_2$GdSbO$_6$ ($A$ = \{Ba, Sr, Ca\}) from 0.4-30 K are depicted in Figure \ref{fig:He3_Gd_HC}. The Debye temperatures $T_D$ were determined from fits to the Debye model (Equation \ref{eq:DebyeModel}) from 6-30 K, Table \ref{table:J1_D_Gd_compounds}. No magnetic ordering is observed in the \emph{fcc} compounds down to 0.4 K. The magnetic entropy of Ba$_2$GdSbO$_6$ and Sr$_2$GdSbO$_6$ at 10 K, relative to that at 0.4 K, is 1.6 J/K/mol$_\textrm{Gd}$, corresponding to only 10\% of that available for $S = 7/2$ Heisenberg spins. In contrast, the site-disordered Ca$_2$GdSbO$_6$ exhibits a sharp $\lambda$-type anomaly at 0.52 K, indicative of a long-range ordering transition. The total entropy contained from 0.4 to 30 K is 12 J/K/mol$_\textrm{Gd}$, or 0.7$R\ln(2S+1)$. This ordering transition may be due to the larger dipolar interaction $D$ in Ca$_2$GdSbO$_6$, Table \ref{table:J1_D_Gd_compounds}, which is $\sim3$ times that of Ba$_2$GdSbO$_6$ and Sr$_2$GdSbO$_6$, and/or different Gd$^{3+}$ ion arrangement. %Heat capacity shows no ordering of fcc compounds, in agreement with frustration; Ca compound orders at 0.52 K, possibly due to a large dipolar interaction

\begin{figure}[t!]
    \centering
    \includegraphics[width=\columnwidth]{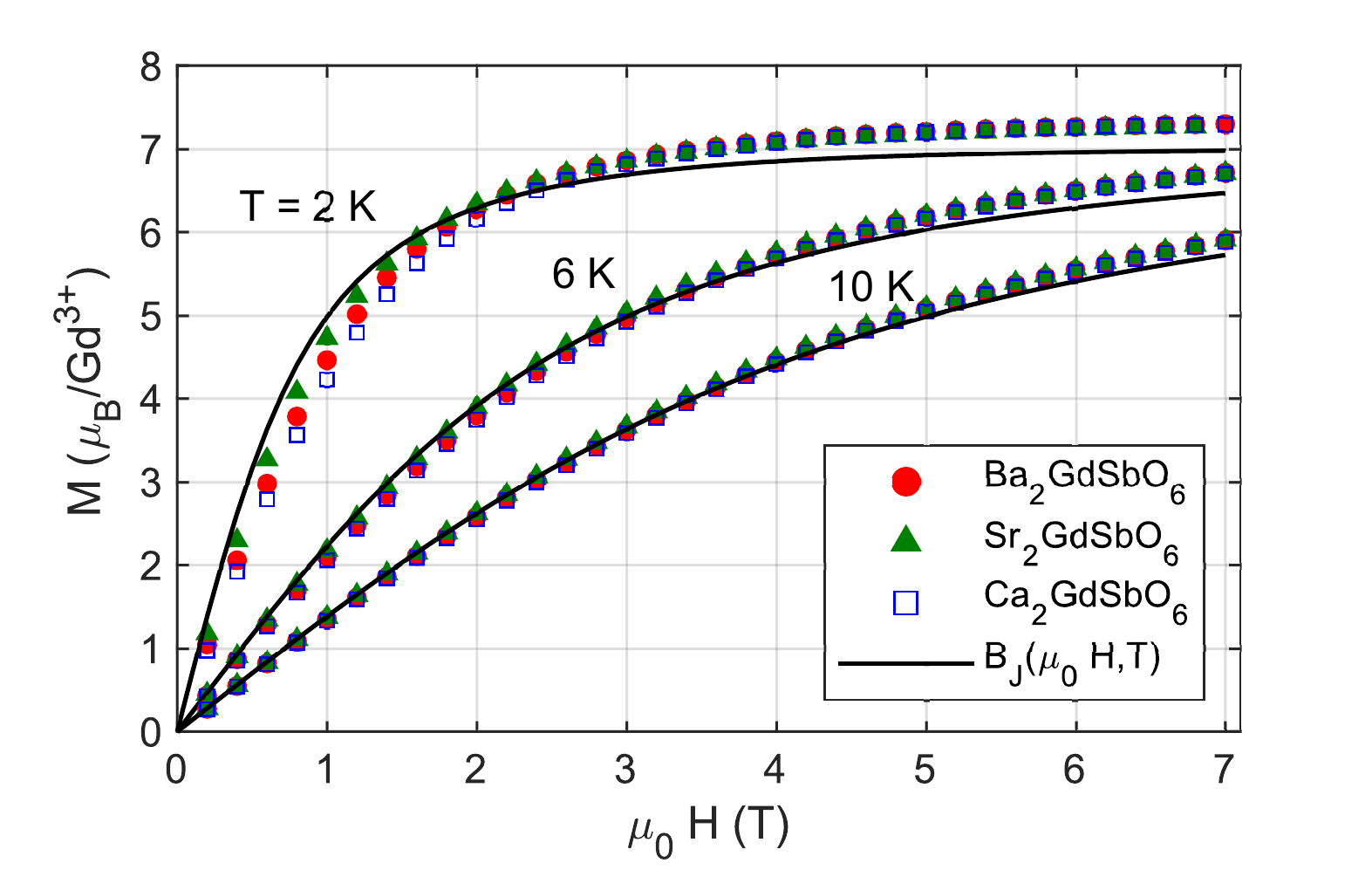}
    \caption{Isothermal magnetization of of $A_2$GdSbO$_6$, $A=$\{Ca,Sr,Ba\} at T = 2, 6, and 10 K compared to the Brillouin function for free $S = 7/2$ spins. Error bars are smaller than the data points.}
    \label{fig:M(H)}
\end{figure}

\begin{figure}[t!]
    \centering
    \includegraphics[width = 0.9\columnwidth]{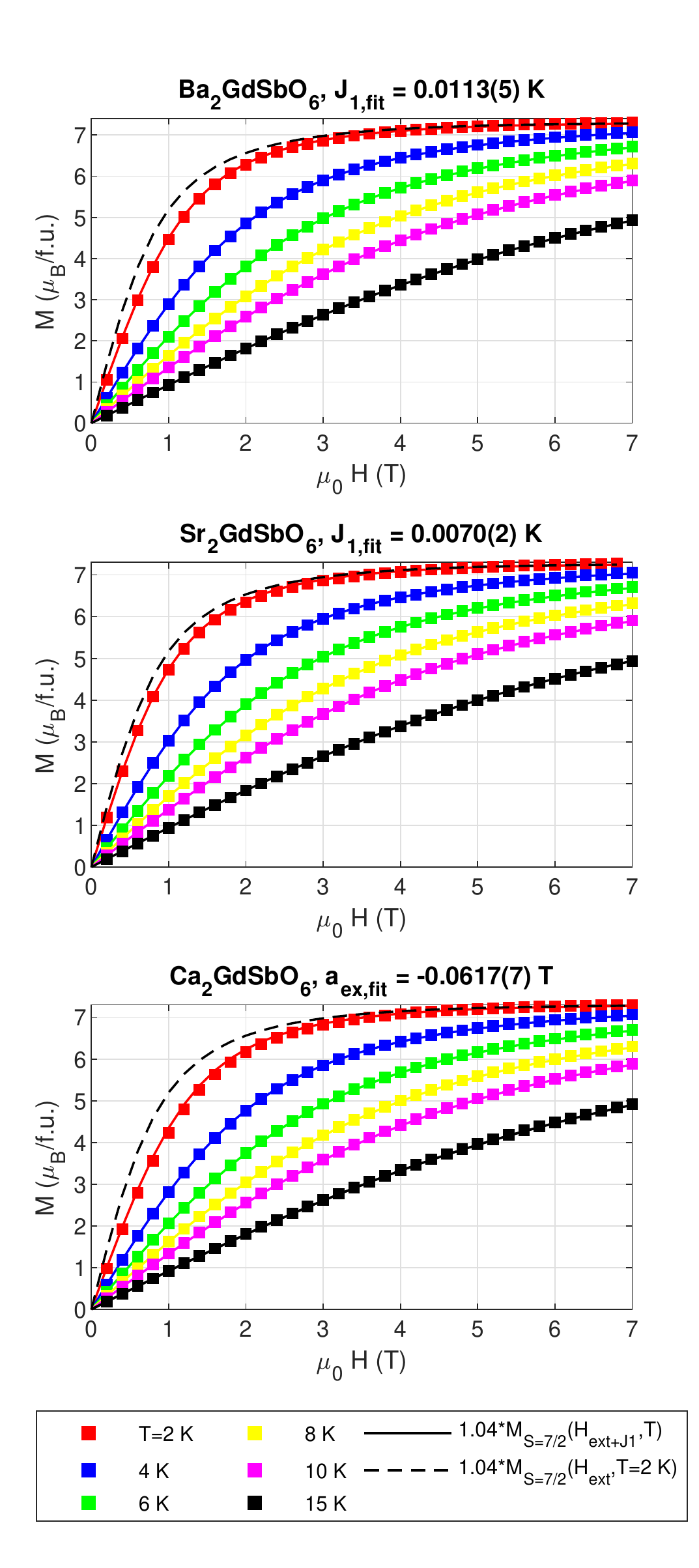}
    \caption{Isothermal magnetization $M$ versus field $\mu_0H$ for $A_2Ln$SbO$_6$ ($A$ =\{Ba, Sr, Ca\}). Measured data are shown as points while theoretical predictions based on a fit of the $nn$ exchange, $J_1$, for each material are shown as solid lines. The prediction of free Heisenberg spins at 2 K is shown as a dashed line. Ca$_2$GdSbO$_6$ was fit to an overall exchange field, $a_{ex}$, with the number of $nn$, z, set to one, due to the presence of site disorder. All error bars are smaller than the data points.}
    \label{fig:MvH_withJ1fits_Gd}
\end{figure}

\begin{figure*}[ht!]
    \centering
    \includegraphics[width=1.98\columnwidth,clip,trim={40 25 0 0}]{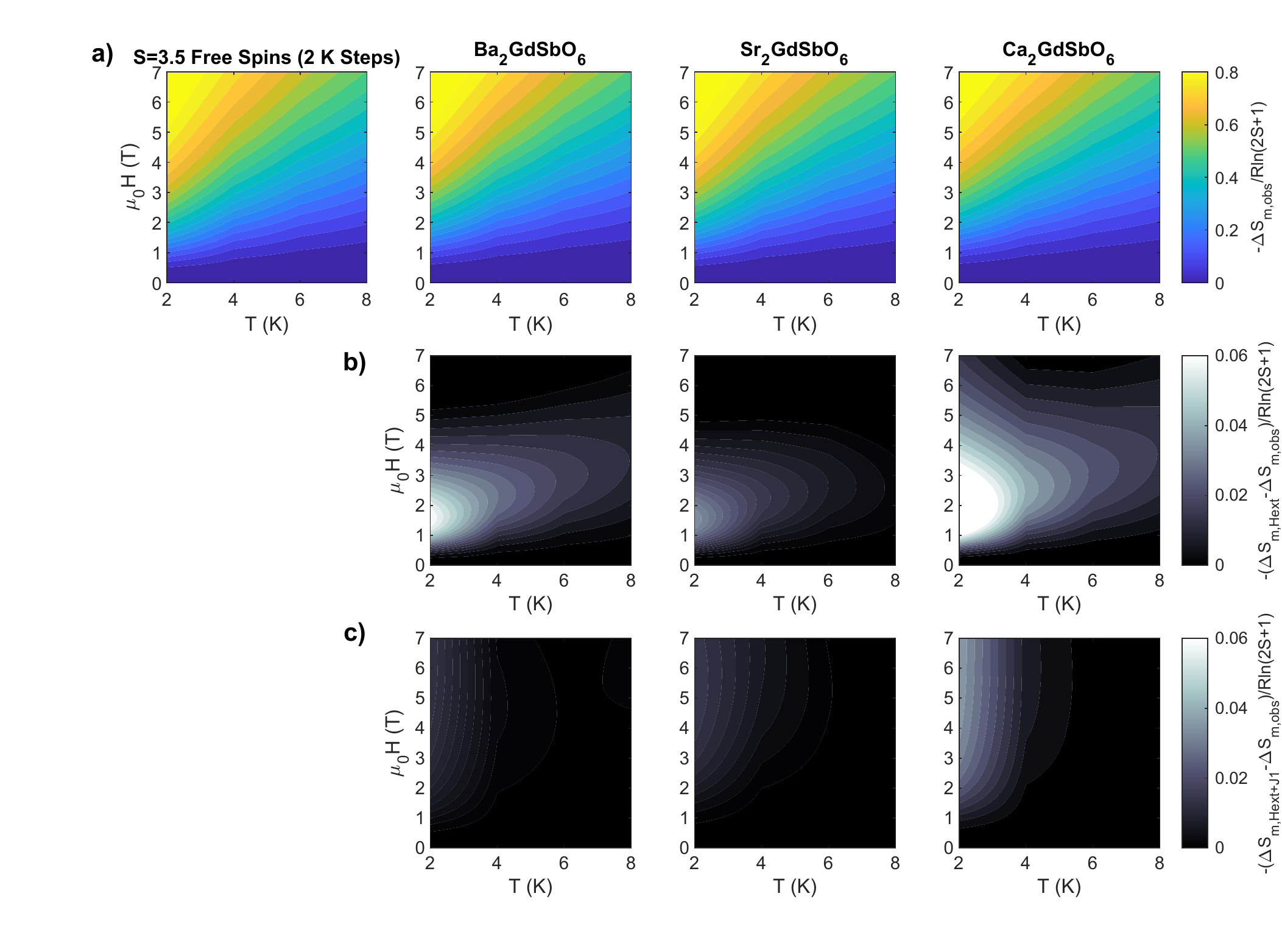}
    \caption{a) Magnetic entropy change $\Delta S_m$ for uncoupled $S = 7/2$ spins determined from the Brillouin function $M_{S=7/2}(H,T)$ with $\Delta T = $2 K and $\Delta H = 0.1$ T steps, compared to the measured $\Delta S_{m,obs}$ for the three $A_2Ln$SbO$_6$ compounds, $A = $\{Ba, Sr, Ca\}. b) Differences between the theoretical entropy change for free-spins, $\Delta S_{m,Hext}$, and that measured for $A_2Ln$SbO$_6$, $\Delta S_{m,obs}$. All compounds exhibit deviations of 0.04$R\ln(2S+1)$ to 0.08$R\ln(2S+1)$ at low (1-2 T), fields possibly indicating the contribution of AFM superexchange. c) Differences between the theoretical magnetic entropy change predicted for the $nn$ exchange field model using the fit $J_1$, $\Delta S_{m,Hext+J1}$, and the measured data, $\Delta S_{m,obs}$ for $A_2Ln$SbO$_6$. Using this $nn$ exchange field model, the deviations are reduced to less than 0.01$R\ln(2S+1)$ for Ba$_2$GdSbO$_6$ and Sr$_2$GdSbO$_6$ and to 0.04$R\ln(2S+1)$ for Ca$_2$GdSbO$_6$.}
    \label{fig:deltaSm_fcc_Gd_withBrillouin}
\end{figure*}

Isothermal magnetization $M(H)$ measurements, shown in Figure \ref{fig:M(H)} and \textcolor{black}{\ref{fig:MvH_withJ1fits_Gd}}, were conducted to measure the magnetocaloric effect, $\Delta S_m(H,T)$. At 2 K, all compounds are saturated by $\mu_0 H = 7$ T at the maximum value for free Heisenberg spins, $g_JJ = 7$ $\mu_B$/Gd$^{3+}$. The measured magnetic entropy change $\Delta S_{m,obs}$ for applied fields of 0.2-7 T at temperatures of 2-\textcolor{black}{8} K is shown in Figure \ref{fig:deltaSm_fcc_Gd_withBrillouin}. The $A_2$GdSbO$_6$ compounds are a high performing set of dense lanthanide oxides, reaching above 90\% of the maximum entropy change (per mol Gd) predicted for uncoupled Heisenberg spins, $R\ln(2S+1)$, in a 7 T field at 2 K. Somewhat surprisingly, the presence of site disorder and magnetic ordering at 0.52 K in [CaGd]$_A$[CaSb]$_B$O$_6$ has only a small effect on the overall magnetocaloric performance in this temperature regime, suggesting that minimal superexchange may play a role in enhancing the magnetocaloric effect in the liquid He regime. %Isothermal magnetization curves show saturation at value expected of free Heisenberg spins --> used these to compute the magnetocaloric effect, finding significant performance compared to other dense Ln oxides

\section{Investigating the Role of Superexchange on the High Magnetocaloric Effect}

To elucidate the origin of the large magnetocaloric effect in these materials, we investigate two models: first, an uncoupled model of $S = 7/2$ Heisenberg spins, and second, a mean-field model that accounts for antiferromagnetic superexchange between Gd$^{3+}$ ions.
%free spin derivation shows small deviations, but can be improved with incorporation of small mean-field exchange term

\begin{table*}[t]
\caption{Fit nearest neighbor exchange $J_1$ for the \emph{fcc} Ba$_2$GdSbO$_6$ and Sr$_2$GdSbO$_6$, and overall exchange field, $a_{ex}z$, for site disordered Ca$_2$GdSbO$_6$ based on Curie-Weiss analysis of the inverse magnetic susceptibility $\chi^{-1}$ (Equation S1 and a mean-field model fit (Equations \ref{eq:Wellm_Hexc} and \ref{eq:Wellm_trans_eq_M}) to the isothermal magnetization curves from 2-20 K. Quoted uncertainties represent a 95 percent confidence interval from the least squares fits.}
\label{table:J1_fits_from_MFT}
\centering
%\begin{ruledtabular}
\begin{tabular}{lcccccc}
\hline
\hline
& & & & & & \\[\dimexpr-\normalbaselineskip+2pt]
 & & \multicolumn{2}{c}{from $\chi^{-1}$ fit} & & \multicolumn{2}{c}{from $M(H)$ fit} \\[0.15cm]
 & & $\Theta$ (K) & $J_1 = \frac{3|\Theta|}{zS(S+1)}$ (K) & & $J_1$ (K) & $R^2$\\[0.15cm]
\hline
& & & & & & \\[\dimexpr-\normalbaselineskip+2pt]
Ba$_2$GdSbO$_6$ & & -0.78(1) & 0.0124(2) & & 0.0113(5) & 1.0000 \\ [0.15cm]
Sr$_2$GdSbO$_6$ & & -0.51(1) & 0.0081(2) & & 0.0070(2) & 1.0000 \\ [0.15cm]
\hline
& & & & & & \\[\dimexpr-\normalbaselineskip+2pt]
 & & $\Theta$ (K) & $z a_{ex}= -\frac{3|\Theta|k_B}{S(S+1)g^2 \mu_B}$ (T) & & $z a_{ex}$ (T) & $R^2$ \\[0.15cm]
%\vspace{1pt}
\hline
& & & & & & \\[\dimexpr-\normalbaselineskip+2pt]
Ca$_2$GdSbO$_6$ & & -0.92(1) & -0.0652(5) & & -0.0617(7) & 0.9999 \\ [0.1cm]
\hline
\hline
\end{tabular}
%\end{ruledtabular}
\end{table*}

\subsection{Uncoupled spin analysis}

Predictions for the theoretical magnetic entropy change $\Delta S_{m,Hext}$ of uncoupled Gd$^{3+}$ ($S$=7/2) spins were computed from the isothermal magnetization curves determined from the Brillouin function and maximum saturation $g_J J$ (Equation S4). M(H,T) curves were evaluated at 2-10 K, with 2 K steps, and from 0-7 T with 0.1 T steps, in accordance with the measured temperatures and fields. Figure \ref{fig:deltaSm_fcc_Gd_withBrillouin} a) and b) demonstrate that the predictions, $\Delta S_{m,Hext}$, for paramagnetic $S$=7/2 spins are remarkably close to the measured entropy changes, $\Delta S_{m,obs}$, for the three $A_2$GdSbO$_6$ compounds. Ba$_2$GdSbO$_6$ and Ca$_2$GdSbO$_6$ exhibit a maximum deviation of 0.09$R\ln(2S+1)$, corresponding to 10\% of the maximum entropy change of 0.9$R\ln(2S+1)$ predicted for free-spins. These deviations occur at low fields (1-3 T) and smaller temperatures ($\sim$2-4 K), in accordance with small antiferromagnetic exchange indicated in the Curie-Weiss analysis. The deviations of $\Delta S_m$ from $\Delta S_{m,S=7/2}$ for Sr$_2$GdSbO$_6$ are lower, only 0.04$R\ln(2S+1)$ and are concentrated at 2 K. None of the measured compounds exceed the magnetic entropy change predicted for free $S = 7/2$ spins; this result is in agreement with the $\sim$1 K Curie-Weiss temperatures of the materials which imply that for the measured 2-10 K temperatures, the materials are paramagnetic.

\subsection{Incorporating an exchange field}

A recent paper on the kagome compound Gd$_3$Mg$_2$Sb$_3$O$_{14}$ showed that a $nn$ exchange field can be used to explain deviations from free-spin behavior below the saturation field \cite{Wellm_GMSO_2020}. Here, we apply this model to characterize the $nn$ exchange $J_1$ in $A_2$GdSbO$_6$ ($A$=\{Ba,Sr,Ca\}) and its role in the isothermal field gradient of the entropy, $(\partial S/\partial H)_T = (\partial M/\partial T)_H$, the determining factor in the magnetocaloric effect.

The model treats antiferromagnetic coupling between $S$=7/2 spins using a mean-field approach so that the net field experienced by a single spin, $S_i$, is composed of the external field $H_{ext}$ and the exchange field $H_{exc}$ due to $z$ nearest neighbors, which scales with the bulk magnetization of the system \cite{Wellm_GMSO_2020}. This mean-field approach is justified for $A_2$GdSbO$_6$ because the Curie-Weiss analysis indicates that the compounds are paramagnetic in the given temperature range (2-22 K) and thus that the role of quantum fluctuations need not be considered.

Since $L = 0$ for Gd$^{3+}$, the exchange constant $J_1$ can be assumed to be isotropic, so that the exchange field is given by:
\begin{equation}
    \vec{H}_{exc} = a_{ex} z M \hat{H}_{ext} = \frac{-J_1}{g^2 \mu_B} z M \hat{H}_{ext},
    \label{eq:Wellm_Hexc}
\end{equation}
where $M$ is the bulk magnetization in units of the Bohr magneton and $a_{ex}$ is the ``field parameter'' in units of magnetic field \cite{Wellm_GMSO_2020}. The bulk magnetization of the system at a given temperature $T$ and external field $H_{ext}$ is given by the roots of the transcendental equation:
\begin{equation}
    f(M) = M - g S B_J(|\vec{H}_{tot}(M)|,T),
    \label{eq:Wellm_trans_eq_M}
\end{equation}
where $B_J$ is the Brillouin function (Equation S4\cite{Wellm_GMSO_2020}.

Using this model, estimates of the nearest neighbor exchange $J_1$ in Ba$_2$GdSbO$_6$ and Sr$_2$GdSbO$_6$ and overall exchange field $a_{ex}$ in Ca$_2$GdSbO$_6$ were found using least squares fits of the observed isothermal magnetization curves $M(H)$ at 2-20 K, Figure \ref{fig:MvH_withJ1fits_Gd}. The free-spin magnetization $M_{S=7/2} = g S B_J (|\vec{H_{ext}}|,T)$ was used as an initial parameter for the mean-field exchange model magnetization (Equation \ref{eq:Wellm_trans_eq_M}). For the \emph{fcc} $A = \{$Ba, Sr\}, the number of $nn$, $z$, was set to 12, while the site disordered Ca version was fit to an overall exchange field, $a_{ex}$, with $z = 1$. All compounds were fit with a scaled fraction of $M_{sat} = g S$, here, $M_{sat} = 1.04 g S \mu_B$, the observed saturated value of the magnetization. 

Table \ref{table:J1_fits_from_MFT} shows the fit $nn$ exchange interaction $J_1$ and overall exchange field $a_{ex}$ for each of the materials, compared to estimates from low field susceptibility measurements.  Overall, there is broad agreement across the two methods. The Curie-Weiss superexchange estimates are slightly larger than from the $M(H)$ curves which could be due to a small contribution from thermally excited states at the higher temperatures fit or due to the fact that only one field was considered in the Curie-Weiss fits.

The role of superexchange in the magnetocaloric effect is examined in Figure \ref{fig:deltaSm_fcc_Gd_withBrillouin} c), which depicts the difference between the mean-field model predicted entropy change, $\Delta S_{m,Hext+J1}$, and the measured entropy change $\Delta S_{m,obs}$ for 2-20 K and 0-7 T. The mean-field model reduces the difference between the predicted and measured magnetic entropy to 1, 2, and 4\% of $R\ln(2S+1)$ for Ba$_2$GdSbO$_6$, Sr$_2$GdSbO$_6$, and Ca$_2$GdSbO$_6$, approximately 1, 2, and 5\% of the max entropy observed. Furthermore, Figure S3 hows that the mean-field model accurately captures the saturation field, and overall magnitude of $(\partial M/\partial T)_H$ at low temperatures, to within the error of the data, compared to the free-spin prediction for both \emph{fcc} compounds. The mean-field prediction for $(\partial M/\partial T)_H$ of Ca$_2$GdSbO$_6$ at 2 K is not in as good of agreement with the measured data likely due to the presence of site disorder, onset of long-range order ($T_0\sim$0.52 K), or a larger dipolar contribution $D \propto 1/R_{nn}^3$. Further experimental validation of these exchange constants could be accomplished by low temperature neutron magnetic diffuse scattering experiments to probe short and long-range correlations between spins. At temperatures of 2 K and above, the mean-field model with antiferromagnetic superexchange thus serves as a good prediction of the observed magnetocaloric effect in $A_2$GdSbO$_6$.

\section{Comparison to Top Performing Gd$^{3+}$ Magnetocaloric Materials}
\label{section:comparison}
\begin{figure}[t!]
    \centering
    \includegraphics[width=0.98\columnwidth]{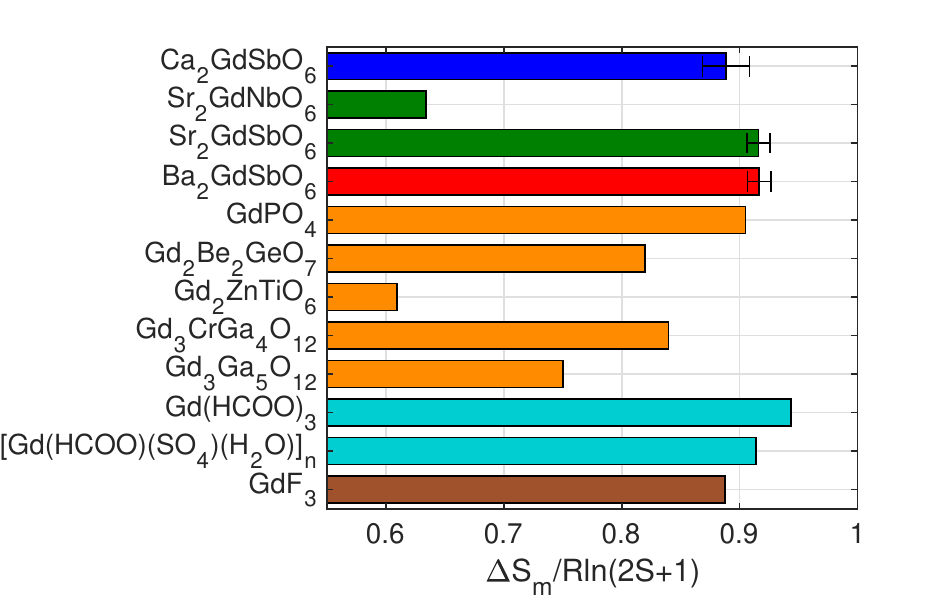}
    \caption{Magnetic entropy change $\Delta S_m$ of $A_2Ln$SbO$_6$ ($A$=\{Ba,Sr,Ca\}) in J/K/mol$_\textrm{Gd}$ compared to top performing magnetocaloric materials from $H_{max} = 7$ T to zero field at 2 K, scaled by $R\ln(2S+1)$ \cite{GdF3_MCE_2015,Gd-formate_2013,GdPO4_MCE_2014,Gd2ZnTiO6_cryogenic_2021,Gd_formatesulfatehybrid_2015,Gd2Be2GeO7_2021,GGG_chromate_2020,Sr2GdNbO6_MCE}. Dense lanthanide oxides are shown in orange, formate-based magnetocalorics in turquoise, ligand-based compounds in brown, and the \emph{fcc} lanthanide oxides in blue, green, and red. Note that the Gd$_2$TiZnO$_6$ value represented is at 2.1 K, as it occurs below the ordering transition. Error bars for $\Delta S_m$ of $A_2$GdSbO$_6$ were estimated using propagation of errors for a mass uncertainty of $\pm 0.1$ mg.}
    \label{fig:bar_graph}
\end{figure}

The magnetic entropy change $\Delta S_m$ attained by the $A_2$GdSbO$_6$ compounds at 2 K and a field of 7 T is compared to other top-performing Gd-based magnetocaloric materials in Figure \ref{fig:bar_graph}. When comparing the entropy change per mol Gd, the \emph{fcc} $A_2$GdSbO$_6$ compounds outperform the top dense oxide magnetocaloric, GdPO$_4$, exhibiting an entropy change of 0.92 $\pm0.1R\ln(2S+1)$, only 0.02$R\ln(2S+1)$ below Gd(HCOO)$_3$ \cite{GdPO4_MCE_2014,Gd-formate_2013}. It should be noted this order of performance would inevitably change when comparing $\Delta S_m$ per unit volume or per unit mass; however, evaluating $\Delta S_m$ per mol Gd highlights the role of superexchange in the magnetocaloric effect. For example, the other top performers Gd(HCOO)$_3$ and GdPO$_4$ have been reported to behave like paramagnets down to 2 K with minimal antiferromagnetic correlations, and GdF$_3$ similarly with minimal ferromagnetic correlations (Table \ref{table:J1_D_Gd_compounds}). Along with the mean-field and free-spin analysis in the preceding sections, these results suggest that minimal coupling between spins plays an important role in maximizing the magnetocaloric effect. 

Table \ref{table:J1_D_Gd_compounds} lists the $nn$ exchange constant $J_1$ and dipolar interaction $D$ in each of the materials estimated from the reported Curie-Weiss temperature $\Theta$ and crystal structure.

\begin{table*}[ht!]
\caption{Curie-Weiss temperature $\Theta$, ordering temperature $T_0$, Debye temperature $T_D$, and corresponding estimates for the mean-field $nn$ exchange $J_1$ and dipolar interaction $D$ for $A_2$GdSbO$_6$ ($A$=\{Ba,Sr,Ca\}) and reported top-performing Gd-based magnetocaloric materials compared to three commonly used paramagnetic salts: ferric ammonium alum (FAS), copper ammonium sulfate (CAS), and copper potassium sulfate (CPS). The mean field $nn$ exchange was calculated using Equation S1 and the dipolar interaction using $D=D_{nn}/S(S+1)$ (with $D_{nn}$ from Equation S2 as in \cite{Paddison_GGG_2015,Paddison_Gd2Sn2O7_2017}.\textsuperscript{\emph{a}}}
\label{table:J1_D_Gd_compounds}
\centering
%\begin{ruledtabular}
\begin{tabular}{lcccccc}
\hline
\hline
 &
$\Theta_{CW}$ (K)&
$J_1$ (K)&
$D$ (K) & $D/J_1$ & $T_0$ (K) & $T_D$ (K) \\ [1pt]
\hline
Ba$_2$GdSbO$_6$ & -0.78(1) & 0.0124(2) & 0.0116 & 0.94 & $<$0.4 & 365 \\ [1pt]
Sr$_2$GdSbO$_6$ & -0.51(1) & 0.0081(2) & 0.0123 & 1.5 & $<$0.4 & 475 \\ [1pt]
Ca$_2$GdSbO$_6$ ($z\approx$ 6) & -0.92(1) & 0.029 & 0.037 & 1.3 & 0.52 & 360 \\ [1pt]
Sr$_2$GdNbO$_6$ \cite{Sr2GdNbO6_MCE} & 3.2 & -0.051 & 0.012 & -0.24 & $\sim2$ & - \\[1pt]
Gd(HCOO)$_3$ \cite{Gd-formate_2013} & -0.3 & 0.0286 & 0.0393 & 1.4 & 0.8 & 168 \\ [1pt]
GdPO$_4$ \cite{GdPO4_MCE_2014} & -0.9 & 0.029 & 0.0362 & 1.3 & 0.8 & 220 \\ [1pt]
GdF$_3$ \cite{GdF3_MCE_2015} & +0.7 & -0.067 & 0.0503 & -0.75 & 1.25 & 284(3) \\ [1pt]
Gd$_3$Ga$_5$O$_{12}$ \cite{Paddison_GGG_2015,Petrenko_GGG_Dipolar_term_2000,GGG_DebyeT_1988} & -2.6(1) & 0.107 & 0.0457 & 0.43 & $\approx 0.14$ & $\approx 500$ \\ [1pt]
Gd$_2$ZnTiO$_6$ \cite{Gd2ZnTiO6_cryogenic_2021} & -4.0 & 0.024 & 0.044 & 1.84 & 2.43 & 156.4 \\ [1pt]
Gd$_2$Be$_2$GeO$_7$ \cite{Gd2Be2GeO7_2021} & -4.09(5) & 0.156(2) & 0.043 & 0.28 & $<$2 & - \\ [1pt]
FAA \cite{Cooke_1949_FAA,FAA_orderingT,FAA_Td} & 0.042 & -0.007 & 0.010 & -1.4 & 0.026 & 80 \\[1pt]
CAS \cite{Cusulfates_nndists, CAS_CPS_Thetas} & 0.010(5) & -0.007(3) & 0.0070 & -1.1 & - & - \\[1pt]
CPS \cite{Cusulfates_nndists, CAS_CPS_Thetas} & 0.016(5) & -0.010(3) & 0.0070 & -0.7 & - & -  \\[1pt]
\hline
\hline
\end{tabular}
%\end{ruledtabular}

\begin{flushleft}
\textsuperscript{\emph{a}} The $nn$ exchange estimated from the $\Theta$ reported for Gd(HCOO)$_3$ and GdF$_3$ was taken to be along the Gd-Gd chains, so that $z$=2, and in the Gd-Gd planes for Gd$_2$Be$_2$GeO$_7$ so that $z$=5. Gd$_2$ZnTiO$_6$ was treated as having 6 $nn$ with an average distance of 3.83 \AA. Ca$_2$GdSbO$_6$ was treated as having $z$=6 with an average distance of 4.063(3) \AA \hspace{1pt}, as described in the text. CAS and CPS were treated as having 6 $nn$ with an average distance of 7.1 \AA \hspace{1pt} as in \cite{Cusulfates_nndists} and FAA as having 2 $nn$ at 6.24 \AA.

\textsuperscript{\emph{b}} The reports of magnetism in Gd$_3$Ga$_5$O$_{12}$ are highly sample dependent.\cite{Schiffer_GGG,GGG_sample_dep}
\end{flushleft}
\end{table*}

The top performing materials, Gd(HCOO)$_3$, and GdF$_3$ all have a $D$/$J_1$ ratio on the order of 1-1.5, indicating that a small dipolar interaction may also improve magnetocaloric performance. Notably $J_1$ for the \emph{fcc} $A_2$GdSbO$_6$ is around 15 mK or less, approximately 0.1-0.5 of the estimated $nn$ exchange in the other materials shown and comparable to common paramagnetic salts, including FAA, CAS, and CPS \cite{Cooke_1949_FAA,CAS_CPS_Thetas}. Aside from Gd$_3$Ga$_5$O$_{12}$, Sr$_2$GdSbO$_6$ and Ba$_2$GdSbO$_6$ have the largest Debye temperatures, indicating the smallest lattice heat capacities, an ideal property in magnetocaloric applications \cite{GdF3_MCE_2015, GdPO4_MCE_2014, Gd-formate_2013}.

The $A_2$GdSbO$_6$ ($A$=\{Ba,Sr,Ca\}) materials investigated in this work provide evidence that minimal superexchange is important in enhancing the magnetocaloric effect in lanthanide oxides. Furthermore, the frustrated \emph{fcc} geometry of $A$ = \{Ba, Sr\} and antiferromagnetic superexchange enable enhanced cooling to at least 400 mK in contrast to some non-frustrated candidates such as GdF$_3$ and Gd(HCOO)$_3$ which are limited to their ordering temperatures of 1.25 and 0.8 K respectively. Although Gd(HCOO)$_3$ may exhibit a better magnetocaloric effect per unit volume or mass, the \emph{fcc} double perovskite structure is more chemically tuneable, and thus allows for  \textcolor{black}{the temperature and magnitude of $\Delta S_m$ to be tuned}. For example, one useful future study would be to investigate partial substitution of Sb$^{5+}$ on the $B$ sites, or $A$ site substitution. For Gd$_3$Ga$_5$O$_{12}$, replacement of a single Ga$^{3+}$ ion with Cr$^{3+}$ improved the entropy change by over 10\% \cite{GGG_chromate_2020}.

The role of the $M^{5+}$ B site ion in the superexchange is highlighted by the recent report of the magnetocaloric effect in Sr$_2$GdNbO$_6$. Sr$_2$GdNbO$_6$ shows differing fundamental magnetic properties (i.e. ferromagnetic interactions), resemblant of $d^0$ versus $d^{10}$ distinctions observed in transition metal oxides \cite{mustonen_spin-liquid-like_2018,mustonen_diamagnetic_2020}. This material exhibits a maximum magnetocaloric effect near its ordering temperature (3 K) for $\mu_0H = 7$ T, -15.5 J/K/mol \cite{Sr2GdNbO6_MCE}, comparable to the performance of Sr$_2$GdSbO$_6$ at 2 K reported in this work.

Our results indicate that changes in the magnetic lattice, such as site disorder in $A$=Ca, do not substantially alter the magnetocaloric effect for the $A_2$GdSbO$_6$ series at $T \geq$2 K. However, disorder does play a role in the magnetic ordering of the compounds, with $A$=Ca exhibiting a transition at 0.52 K and $A$=\{Ba,Sr\} remaining disordered down to 0.4 K. Future low-temperature heat capacity in field, $\mu$-SR, and/or low-temperature neutron diffraction using isotopically-enriched samples will be important in understanding how disorder affects the low temperature magnetic behavior. \textcolor{black}{
%Disorder has recently been shown to play a role in the MCE observed in $A$GdS$_2$, $A$=\{Li,Na\}, with a significant enhancement of the MCE in disordered NaGdS$_2$ compared to cation ordered LiGdS$_2$. \cite{AGdS2_MCE}
Disorder has recently been shown to play a role in the magnetocaloric effect observed in $A$GdS$_2$, A=\{Li,Na\}, with a significant enhancement of the magnetocaloric effect observed in ordered NaGdS$_2$ compared to cation disordered LiGdS$_2$. This is rationalised by differences in the exchange interaction and the onset of ordering at higher temperatures in LiGdS$_2$. At high temperatures, $T >$ 2 K, a similar effect is not observed in the $A_2$GdSbO$_6$ double perovskites but may result in significant differences in the magnetocaloric effect closer to the ordering temperature in Ca$_2$GdSbO$_6$.\cite{AGdS2_MCE}}

The \emph{fcc} materials presented here, Ba$_2$GdSbO$_6$ and Sr$_2$GdSbO$_6$, are likely able to cool below the industry standard, Gd$_3$Ga$_5$O$_{12}$ (which has a lower cooling limit of $\sim 0.8$ K due to spin-spin correlations \cite{GGG_limits_MCE}), into the temperature regime of paramagnetic salts ($\sim 100$ mK or less \cite{wikus_magnetocaloric_2014}) based on their minimal superexchange. This presents a possible significant advancement as the frustrating lattice should have a better per unit volume magnetic entropy change than a Gd$^{3+}$-based paramagnetic salt.

\section{Conclusion}
We synthesized three frustrated lanthanide oxides $A_2$GdSbO$_6$ ($A = $\{Ba,Sr,Ca\}) and characterized their structural and magnetic properties through x-ray powder diffraction and bulk magnetic measurements. The frustrated \emph{fcc} lattice and small ($J_1 \sim 10$ mK) antiferromagnetic superexchange of Ba$_2$GdSbO$_6$ and Sr$_2$GdSbO$_6$ prevents magnetic ordering down to 0.4 K. In contrast, Ca$_2$GdSbO$_6$ is found to be site-disordered, with all Gd$^{3+}$ ions lying on the $A$ sites and an antiferromagnetic ordering transition at 0.52 K.

Intriguingly, all three materials make promising magnetocaloric candidates in the liquid He regime (2-20 K), achieving up to 92(1)\% of the ideal magnetic entropy change $R\ln{(2S+1)}$ in an applied field of up to 7 T. The comparable, high magnetocaloric performance ($\Delta S_m$ = 0.88(2)$R\ln{(2S+1)}$) of the site-disordered compound Ca$_2$GdSbO$_6$ suggests that the magnetocaloric effect is governed by primarily free spin behavior at these temperatures. We demonstrate that the measured magnetocaloric effect of the frustrated Ba$_2$GdSbO$_6$ and Sr$_2$GdSbO$_6$ can be predicted to within experimental uncertainty using a mean-field model with a fit $nn$ superexchange constant, $J_1$. These results suggest that future top-performing Gd-based magnetocaloric materials should search for a balance between minimal superexchange between magnetic ions and frustration to suppress the magnetic ordering temperature. The tunability of the double perovskites via chemical substitution makes the \emph{fcc} lanthanide oxides a promising set of materials for magnetic refrigeration.
\textcolor{black}{
\section{Supporting Information}
The Supporting Information is available free of charge at;}

\textcolor{black}{additional magnetic characterisation, free Heisenberg spin model, $M(H)$ mean field exchange model. (PDF) 
\\Data associated with this publication can be found at; }

\begin{acknowledgement}

This work was supported through the EP/T028580/1 EPSRC grant and the Winton Programme for the Physics of Sustainability. The work of E.C.K. was supported by a Churchill Scholarship from the Winston Churchill Foundation of the United States. N.D.K. gratefully acknowledges the EPSRC for the provision of a PhD Studentship (EP/R513180/1). Magnetic measurements at Cambridge were made on the EPSRC Advanced Characterization Suite EP/M0005/24/1. We acknowledge Diamond Light Source for time on I11 under BAG proposal CY28349 and Dr. Sarah Day for collecting the data. The authors thank Dr. Jiasheng Chen for his assistance in the low temperature 3-He heat capacity measurements and Dr. Joshua Bocarsly for useful discussions regarding crystal refinements of the I11 data from the Diamond Light Source.
\end{acknowledgement}

\textcolor{black}{The authors declare no competing financial interest.}

%\nocite{*} %prints out all references in .bib regardless of whether cited or not

\bibliography{bibfile}% Produces the bibliography via BibTeX.

\newpage\hbox{}\thispagestyle{empty}\newpage

\begin{figure}
    \centering
    \includegraphics[width=8.5cm]{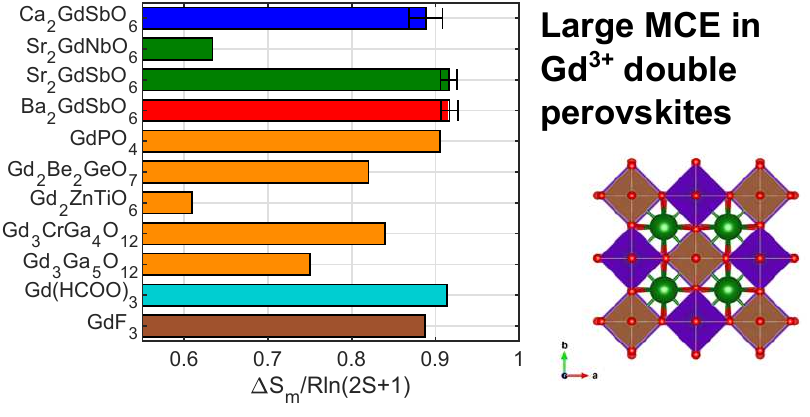}
    \caption{\textcolor{black}{Table of Contents Figure}.}
    \label{fig:small_TOC}
\end{figure}

\end{document}